\documentclass{pasa}%

\title[Polarized sub-second spikes in UV Ceti flare]{Discovery of the subsecond
  linearly polarized spikes of synchrotron origin in the UV Ceti giant optical
  flare}
\author[G. Beskin et al.]{
  G. Beskin,$^{1,2}$\thanks{E-mail: beskin@sao.ru}
  S. Karpov,$^{1,2}$
  V. Plokhotnichenko,$^{1}$
  A. Stepanov$^{3}$
  and Yu. Tsap$^{3}$
\\
\affil{$^{1}$Special Astrophysical Observatory, Nizhnij Arkhyz, Karachaevo-Cherkessia
369167, Russia}
\affil{$^{2}$Kazan (Volga region) Federal University, Kazan, 420008, Russia}
\affil{$^{3}$Pulkovo Observatory of Russian Academy of Sciences, Saint-Petersburg,
196140, Russia}}

\jid{PASA}
\doi{10.1017/pas.\the\year.xxx}
\jyear{\the\year}

\usepackage[authoryear]{natbib}
\bibpunct{(}{)}{;}{a}{}{,}
\usepackage{aas_macros}
\usepackage{hyperref}
\hypersetup{colorlinks,citecolor=blue,linkcolor=blue,urlcolor=blue}

\usepackage{textcomp}

\def\icarus{\ref@jnl{Icarus}}   
\def\rmxaa{\ref@jnl{Rev.~Mex.~AA.}}   

\begin{document}%
\begin{abstract}

  During our optical monitoring of UV Ceti, iconic late-type flaring star, with
  high temporal resolution using the Russian 6-m telescope in 2008 we detected
  a giant flare with the amplitude of about 3 magnitudes in $U$-band. Near flare
  maximum more than a dozen of spike bursts have been discovered with
  triangular shapes and durations from 0.6 to 1.2 s and maximal luminosities in
  the range $(1.5-8)\times10^{27}$ erg~s$^{-1}$. For the half of these events the
  linear polarization exceeds 35\% with significance better than $5\sigma$.  We
  argue that these events are synchrotron emission of electron streams with the
  energies of several hundred MeV moving in the 
  magnetic field of about
  1.4 kG. Emission from such ultrarelativistic (with energies far exceeding 10
  MeV) particles is being routinely observed in solar flares, but has never
  been detected from UV Ceti type stars.

  This is the first ever detection of linearly polarized optical light from the UV
  Ceti-type stars which indicates that at least some fraction of the flaring
  events on these stars are powered by a non-thermal synchrotron emission
  mechanism.

\end{abstract}
\maketitle%

\section{Introduction}

Presently there are no doubts that the flaring activity of the Sun and flaring stars,
particularly those of UV~Cet-type, is of a common origin
\citep{gershberg_2015}. Flaring events are caused by the release of energy
stored in coronal magnetic fields. Therewith up to 10--50\% of the magnetic
energy is converted to the kinetic energy of accelerated electrons and ions in
solar flares \citep{lin_1976,miller_1997,holman_2003}. They are partly ejected
away from the star and partly move along coronal magnetic loops, heating dense
regions of the stellar atmosphere and generating the flare-like emission in a
wide range of frequencies, from radio to gamma-rays \citep{priest_2000,benz_2010}.

On the Sun, these initial electrons are accelerated up to hundreds of MeV -- 1
GeV, as observations of gamma-ray emission of flares suggest (see, for example,
\citet{kanbach_1993, ramaty_1994, aschwanden_2006}). Flaring stars also display
some manifestations of energetic non-thermal particles with Lorentz factors
$1 < \gamma \le 10$ in the form of bursts of gyrosynchrotron radio emission
\citep{bastian_1990, gudel_1996, gudel_2002} correlated with variations of soft
X-ray emission of the chromospheric and photospheric plasma heated by these
particles \citep{gudel_2002, smith_2005, benz_2010}, but the presence of
ultrarelativistic electrons with $\gamma>10$ has never been revealed in these
studies. 
On the other hand, the detection of such particles has been reported as a
result of radio observations of T~Tau stars in binary systems
\citep{massi_2006,salter_2010}.

If such energetic particles exists in UV Ceti like stars too, they
may produce optical synchrotron emission
while moving in the magnetic fields of coronal loops
\citep{gordon_1954}.
Due to discrete nature of particle acceleration driven by the magnetic field
reconnection in a stellar corona, theis emission may be in the form of separate
subsecond spikes.  Similar events, seen on the Sun in X-ray and gamma-ray
bands, are generated by the bremsstrahlung emission of electron beams in the
chromosphere, and have typical durations of 0.05 to 1 seconds
\citep{kiplinger_1983, aschwanden_1995, cheng_2012}
\footnote{Even shorter radio spikes, also quite common phenomena in solar and
  stellar flares \citep{fleishman_1998,osten_2006}, are of coherent nature.}
On the other hand, optical spike bursts with sub-second durations in quiet
states of flaring stars have been detected only in a few cases -- a bit more
than dozen of flashes seen in different years and with different instruments
while observing EV Lac, BY Dra, V577 Mon and CN Leo
\citep{zhilyaev_1995, robinson_1995, tovmassian_1997, zhilyaev_1998} and
lacking proper interpretations. A number of small amplitude flares with
durations of a few seconds have recently been reported in UV Ceti \citep{schmitt_2016}.

As the timescale of synchrotron spikes is close to the one of thermal emission
mechanism for optical flashes, which may be down to $0.1-1$ s
\citep{katsova_1986,shvartsman_1988,katsova_2001}, an additional criterion is
necessary to distinguish between these mechanisms -- the presence of linear
polarization, which is a characteristic feature of synchrotron emission
\citep{ginzburg_1965,rybicki_lightman_2004}. Therefore, in order to confidently
detect such synchrotron emission from a flaring star, one has to perform its
regular photo-polarimetric monitoring with high temporal resolution using large
telescope.

Numerous attempts to detect the polarization in
flares of YZ CMi, AD Leo, EV Lac, YY Gem, observed with various telescopes
\citep{karpen_1977, eritsyan_1978, tuominen_1989} were also unsuccessful. Most
reliable data have been acquired during EV Lac study using 1.25 m telescope of
the Crimean Astrophysical Observatory \citep{alekseev_1994}. The upper limits
for polarization degree in strong flares have been placed on 2\% level on 10
s timescale, and 1\% level -- on a 50 s one.

In 1982-1985 we performed a regular photometric monitoring of UV Ceti, CN
Leo, V 577 Mon and Wolf 424 flaring stars using the Russian 6-m telescope with 1
microsecond temporal resolution. More than hundred flashes have been detected
in U band, and the upper limits on the amplitude of intensity variations, both
during the flares and outside them, on time scales from 1 $\mu$s to 1 s have been
placed on the level of 20\% to 0.5\% \citep{beskin_1988a, beskin_1988b}, correspondingly. The
shortest details detected were the rising fronts of four flares of these stars
with durations from 0.3 to 0.8 seconds; decay times of these flares were 1 to 3
seconds (Shvartsman et al., 1988). These data, along with the statistical
properties of all temporal characteristics, have led to conclusion that the
flares, even the shortest ones, may be explained by the gas-dynamic model
\citep{beskin_1988b, katsova_2001, gershberg_2015}.

In order to reliably detect the polarized non-thermal emission from UV Ceti
type stars, since 2008 we started the new set of observations of these
objects with the Russian 6-m telescope using the panoramic photo-polarimeter. In this
study we report the detection of a giant, with amplitude of nearly 3 magnitudes,
flare of dMe-dwarf UV Ceti during this monitoring. Near its maximum, we
discovered more than a dozen of spike bursts with the duration of 0.6 -- 1.2 s,
with linear polarization exceeding 35\% -- 40\% for the majority of them. We argue that
these events were generated by synchrotron emission of electrons with the
energies of several hundred MeV moving in magnetic fields with the strength of
about 1.4 kG, and therefore are the first ever evidence for the presence
of ultrarelativistic electrons in the flares of UV Cet-type stars.

The paper is organized as follows. Section 2 describes the equipment and methods
used for the high temporal resolution photopolarimetric observations of flaring
stars with the Russian 6-m telescope. In Section 3, the observational
characteristics of detected spike bursts are described, and in Section 4 the
physical conditions necessary for their generation are analyzed. Section 4 also
contains the discussion of ultrarelativistic electrons production mechanisms in
UV Ceti like stars. Section 5 presents the brief summary of our results.

\section{Observations and data reduction}

Systematic monitoring of flaring stars with one-microsecond temporal resolution
with the Russian 6-m telescope is ongoing since 2008, using the panoramic
photospectropolarimeter in its various configurations
\citep{plokhotnichenko_2009a}. The detector is the microchannel plate based
position-sensitive photon counter with 4-electrode cathode \citep{debur_2003},
and the high speed data acquisition is performed with the dedicated
``Quantochron 4-48'' card plugged into the PC which encodes and stores the
coordinates and times of arrival of detected photons with 1 $\mu$s accuracy
\citep{plokhotnichenko_2009b}. The data acquired are the photon lists which may
be later arbitrarily binned for light curve and image analysis on various time
scales.

In 2008 the monitoring has been performed in $U$ photometric band using the
Wollaston prism as a polarizer. Field of view of the instrument was about 20
arcseconds and angular resolution -- about 2 arcseconds, with no stars except
for unseparable UV/BL Ceti pair seen. On the photocathode of the detector the
Wollaston prism formed two images of an object with orthogonal orientations of
polarization plane. The intensities of these two images are
\citep{shurkliff_1962, snik_2013}:
\begin{equation}
  I_0 = \frac12 (I + Q)
\end{equation}
\begin{equation}
  I_{90} = \frac12 (I - Q)
\end{equation}

These quantities allow one to determine two of the four Stokes parameters

\begin{equation}
  I = I_0 + I_{90}
\end{equation}
\begin{equation}
  Q = I_0 - I_{90}
\end{equation}

The degree of linear polarization is a combination of $I$, $Q$ and $U$ Stokes
parameters \citep{snik_2013}
\begin{equation}
  P = \frac{1}{I}\sqrt{Q^2 + U^2} \ \ \ .
\end{equation}
We know only the first two, and may therefore place the lower limit on the
degree of linear polarization, as $P \ge |Q| / I$ for all $U$ values.

That is, if we detect any significant deviation of $Q/I$ from zero, then the
lower limit on the degree of linear polarization is
\begin{equation}
  P_{\rm low} = |Q|/I = \frac{|I_0-I_{90}|}{I_0+I_{90}}\label{eq_lower_limit}
\end{equation}

The giant flare of UV Ceti, with nearly $3^{\rm m}$ amplitude and fast ($\sim10$ seconds, as
compared to 30 minutes overall duration) initial rise, has been detected on Dec
28, 2008 at 15:27:02 UT.

\begin{figure}
  {\centering \resizebox*{1.0\columnwidth}{!}{
      \includegraphics[angle=270]{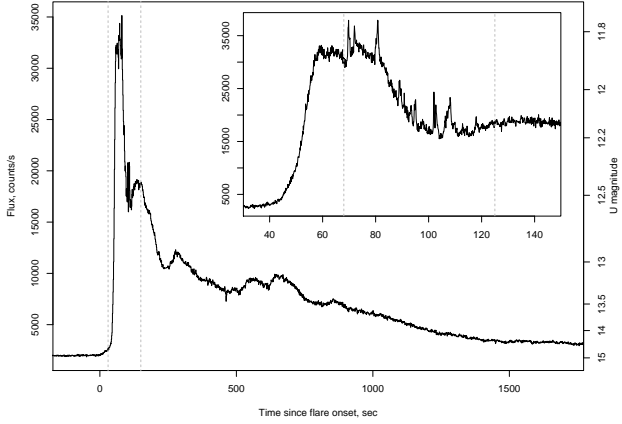}
    }}
  \caption{Light curve of UV Ceti flare in U band with 0.1 s temporal resolution,
    obtained with Russian 6-m telescope on Dec 28, 2008. The inlet shows the region marked with dashed
    lines in the main plot. In turn, the region marked with vertical dashed lines
    in the inlet, containing all the spike bursts, is shown in Figure~\ref{fig2}.
    \label{fig1}}
\end{figure}

The light curve, computed as a sum of background subtracted intensities in two
images, is shown in Figure~\ref{fig1}. As we did not have any other star in the
field of view, to calibrate the flux we associated the average pre-flare flux
(-100 s to 0 s) level with $m_U=14.86$ magnitude of UV / BL Ceti pair
\citep{eason_1992}, and assumed the distance $d=2.68$ pc. The inlet of
Figure~\ref{fig1} also shows the two-minute interval of maximal intensity with
0.1 s temporal resolution. During this interval, more than a dozen of spike
bursts with 6--50\% relative amplitudes and durations not exceeding (except for
two cases) 1.2 s are clearly seen. Their analysis will be performed in the next
section.

As we are interested in polarization variations during the flare, and to
accommodate for the instrumental, atmospheric and interstellar polarization, we
computed the mean value of $k = I_0/I_{90}$ over the 100 seconds long pre-flare
interval, $\left<k\right> = 1.23$, and scaled the $I_{90}$ using this coefficient. This way
normalized Stokes $Q/I$ parameter has zero mean over that interval.

\begin{figure}
  {\centering \resizebox*{1.0\columnwidth}{!}{
      \includegraphics[angle=270]{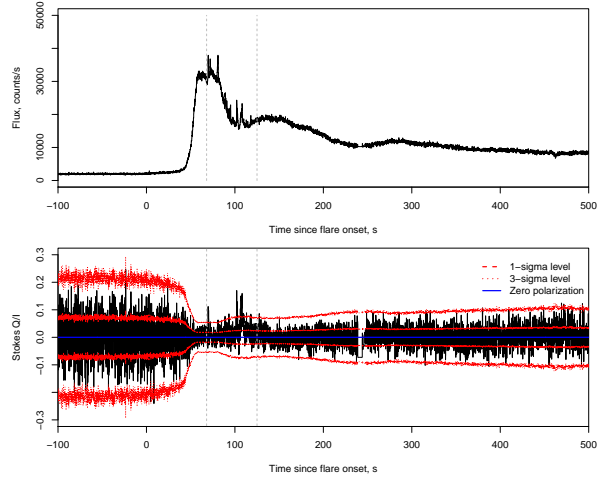}
    }}
  \caption{
    The behaviour of normalized Stokes $Q/I$ parameter during the pre-flare and
    flare main part intervals. Levels of $\sigma$ and $3\sigma$, estimated
    assuming Poissonian statistics, are shown. Vertical lines mark the
    interval where spike bursts are detected. Some of them show significant
    polarization, while all other intervals of the light curve (upper panel) do not.
    \label{fig4}}
\end{figure}

Figure~\ref{fig4} shows the Stokes $Q/I$ over the pre-flare and main part of
the flare interval (the gap from 240 s to 250 s corresponds to the restart of
data acquisition system where the information has been lost). No significant
deviations from zero may be seen, except for the short events coinciding with
the spike bursts detected in the light curve.

\section{Analysis of spikes properties}

\subsection{Durations and shapes}
\label{sec_durations}

\begin{figure}
  {\centering \resizebox*{1.0\columnwidth}{!}{
      \includegraphics[angle=270]{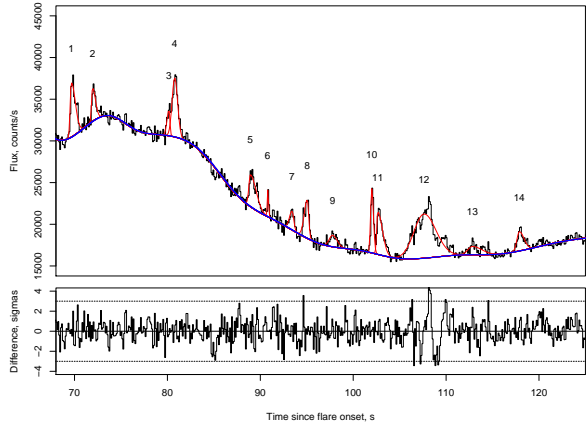}
    }}
  \caption{The fragment of UV Ceti flare light curve containing 14 spike bursts,
    whose shapes are approximated with splitted Gaussian profiles (upper panel),
    approximated with the smooth spline for the background flare and splitted
    Gaussians for the spikes. Lower panel shows the normalized residuals,
    normally distributed everywhere except for the 10-s region around complex
    peak \textnumero 12.
    \label{fig2}}
\end{figure}

Figure~\ref{fig2} shows the region of the light curve with clearly seen flaring
activity, where 14 spike bursts have been detected. Their rise times are 0.2 --
0.5 seconds, and the shapes are nearly symmetric (see upper panel in Figure~\ref{fig2}).

We approximated the slowly-changing flare background with manually adjusted
smooth spline, and the spike shapes with splitted Gaussians, having different
rising and fading characteristic times:
\begin{equation}
  I(t) = \left\{
    \begin{array}{rl}
      A \exp{\left(\frac{\ln{2} \cdot (t-t_0)^2}{S_1^2}\right)} , &  t < t_0 \\
      A \exp{\left(\frac{\ln{2} \cdot (t-t_0)^2}{S_2^2}\right)} , &  t > t_0 \\
    \end{array}
      \right.
  \label{eq_spike_shape}
\end{equation}
where $A$ is the peak intensity, $t_0$ -- peak time and $S$ is the half width at half
maximum -- the characteristic time of a twofold change of the intensity. The
approximation is shown in Figure~\ref{fig2}, and the fit parameters are listed in Table
1. The residuals, shown in lower panel of Figure~\ref{fig2}, are distributed normally --
Shapiro-Wilk normality test can not reject their normality
with $p$-value better than 0.3 (except for 10 second long interval around spike
\textnumero 12, which has very complex shape, and most probably consists of several
blended sub-spikes, which can't be easily separated), and therefore the
approximation is quite optimal.

\begin{table*}
  \caption{Parameters of the light curve spikes. The spikes have been fitted with splitted Gaussian profiles.
  Here $t_0$ -- peak time, $A$ -- peak amplitude, FWHM -- peak full width at half
  maximum, $S_1$ and $S_2$ -- half widths at half maximum of the rising and
  fading fronts, $L$ -- peak
  luminosity in $U$-band.
  Also, $A_0$ and $A_{90}$ represent the amplitudes of the splitted Gaussian profiles
  fitted to $I_0(t)$ and $I_{90}(t)$ intensities, correspondingly, and $\left<P_{\rm
    low,int}\right>$ is the mean intrinsic polarization of the spikes computed
  according to Eq.~\ref{eq_pint}.
  \label{table1}}
\footnotesize
\begin{center}
\begin{tabular}{lccccccccc}
    \hline
    \hline
    \textnumero & $t_0$ & $A$ & FWHM & $S_1$ & $S_2$ & $L$ & $A_0$ & ${A_{90}}^*$ &
    $\left<P_{\rm low,int}\right>$ \\
    & s & cts~s$^{-1}$ & s & s & s & $10^{27}$ erg~s$^{-1}$ & cts~s$^{-1}$ &
    cts~s$^{-1}$ & \\
    \hline
1
& 69.65 $\pm$ 0.04
& 6507 $\pm$ 390
& 0.65
& 0.14 $\pm$ 0.04
& 0.51 $\pm$ 0.05
& 7.28
& 2043 $\pm$ 166
& 5460 $\pm$ 228
& 0.45 $\pm$ 0.05
\\

2
& 71.93 $\pm$ 0.06
& 3957 $\pm$ 447
& 0.49
& 0.12 $\pm$ 0.06
& 0.37 $\pm$ 0.08
& 4.43
& 1166 $\pm$ 190
& 3377 $\pm$ 261
& 0.49 $\pm$ 0.11
\\

3
& 80.17 $\pm$ 0.07
& 3077 $\pm$ 690
& 0.37
& 0.25 $\pm$ 0.09
& 0.12 $\pm$ 0.08
& 3.44
& 1524 $\pm$ 218
& 1858 $\pm$ 298
& 0.10 $\pm$ 0.12
\\

4
& 80.78 $\pm$ 0.06
& 7092 $\pm$ 413
& 0.69
& 0.33 $\pm$ 0.09
& 0.37 $\pm$ 0.05
& 7.93
& 3045 $\pm$ 161
& 4936 $\pm$ 220
& 0.24 $\pm$ 0.04
\\

5
& 88.94 $\pm$ 0.07
& 3878 $\pm$ 273
& 0.92
& 0.20 $\pm$ 0.07
& 0.73 $\pm$ 0.09
& 4.33
& 1435 $\pm$ 138
& 2991 $\pm$ 190
& 0.35 $\pm$ 0.07
\\

6
& 90.79 $\pm$ 0.15
& 3253 $\pm$ 1044
& 0.16
& 0.05 $\pm$ 0.10
& 0.11 $\pm$ 0.12
& 3.64
& 1827 $\pm$ 326
& 1712 $\pm$ 447
& 0.03 $\pm$ 0.15
\\

7
& 93.42 $\pm$ 0.09
& 2269 $\pm$ 326
& 0.55
& 0.33 $\pm$ 0.11
& 0.21 $\pm$ 0.10
& 2.53
& 723 $\pm$ 179
& 1869 $\pm$ 245
& 0.44 $\pm$ 0.17
\\

8
& 95.05 $\pm$ 0.05
& 4610 $\pm$ 327
& 0.56
& 0.38 $\pm$ 0.05
& 0.18 $\pm$ 0.05
& 5.16
& 2636 $\pm$ 176
& 2410 $\pm$ 241
& 0.04 $\pm$ 0.06
\\

9
& 97.78 $\pm$ 0.20
& 1393 $\pm$ 224
& 1.00
& 0.42 $\pm$ 0.21
& 0.58 $\pm$ 0.22
& 1.56
& 586 $\pm$ 132
& 968 $\pm$ 182
& 0.25 $\pm$ 0.18
\\

10
& 102.00 $\pm$ 0.02
& 7718 $\pm$ 428
& 0.36
& 0.15 $\pm$ 0.02
& 0.21 $\pm$ 0.03
& 8.64
& 2282 $\pm$ 220
& 6676 $\pm$ 302
& 0.49 $\pm$ 0.06
\\

11
& 102.65 $\pm$ 0.04
& 5118 $\pm$ 281
& 0.72
& 0.14 $\pm$ 0.04
& 0.59 $\pm$ 0.05
& 5.73
& 1864 $\pm$ 155
& 3993 $\pm$ 213
& 0.36 $\pm$ 0.06
\\

12
& 107.61 $\pm$ 0.09
& 5298 $\pm$ 141
& 2.81
& 1.28 $\pm$ 0.09
& 1.52 $\pm$ 0.10
& 5.93
& 1717 $\pm$ 80
& 4463 $\pm$ 110
& 0.44 $\pm$ 0.03
\\

13
& 112.84 $\pm$ 0.34
& 995 $\pm$ 162
& 1.78
& 0.59 $\pm$ 0.35
& 1.19 $\pm$ 0.39
& 1.11
& 293 $\pm$ 100
& 855 $\pm$ 137
& 0.49 $\pm$ 0.22
\\

14
& 117.88 $\pm$ 0.11
& 2306 $\pm$ 240
& 0.89
& 0.31 $\pm$ 0.11
& 0.59 $\pm$ 0.12
& 2.58
& 764 $\pm$ 140
& 1880 $\pm$ 192
& 0.42 $\pm$ 0.13
\\
\hline
\hline
\end{tabular}
\end{center}
\tabnote{$^*$ To account for instrumental polarization,
  $I_{90}$ has been scaled by a constant coefficient $\left<k\right>=1.23$, and
  therefore the sum of $A_0$ and $A_{90}$ is not equal to $A$.}
\end{table*}

To test the randomness of spikes' peak times we performed Kolmogorov-Smirnov
test on $t_0$ values from Table~\ref{table1}, which gave $p$-value of 0.7,
which confirms that the distribution of spikes in time is uniform.

\begin{figure}
  {\centering \resizebox*{1.0\columnwidth}{!}{
      \includegraphics[angle=270]{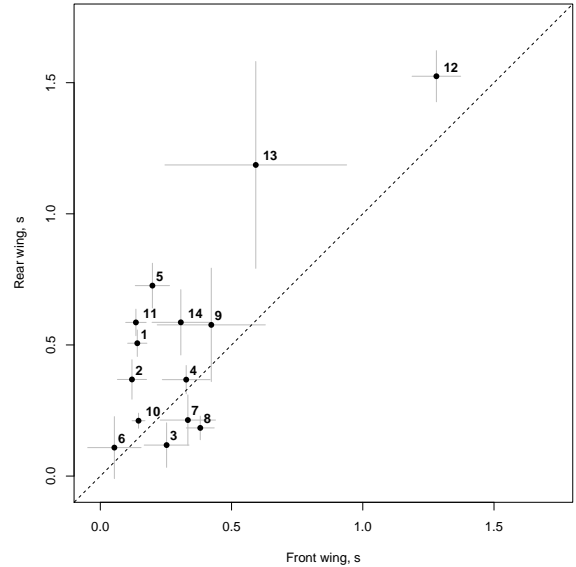}
    }}
  \caption{Comparison of the spikes' rise and fall times. Total durations of 12
    events are less than a second; for 6 events rise and fall times are nearly
    equal, while for 6 others -- the latters are 2--2.5 times greater. Durations of
    two spikes (\textnumero12 and \textnumero13) are 2 -- 3 seconds -- most
    probably, they consist of several overlapping events.
    \label{fig3}}
\end{figure}

Figure~\ref{fig3} displays the spikes' rise and fading times with corresponding fit
errors. If we exclude complex spikes \textnumero 12 and \textnumero 13, then the rise and fading
times are uncorrelated (with Pearson correlation coefficient $r=0.08$) and follow
the same distribution (with Kolmogorov-Smirnov test $p$-value $p=0.26$).  Their
mean values are
$\left<S1\right> = (0.24 \pm 0.04)$ s, $\left<S2\right> = (0.38 \pm 0.06)$ s.
Therefore, the spikes are nearly symmetric and have triangular form.

\subsection{Polarization of spikes}

\begin{figure*}
  {\centering \resizebox*{1.8\columnwidth}{!}{
      \includegraphics[angle=270]{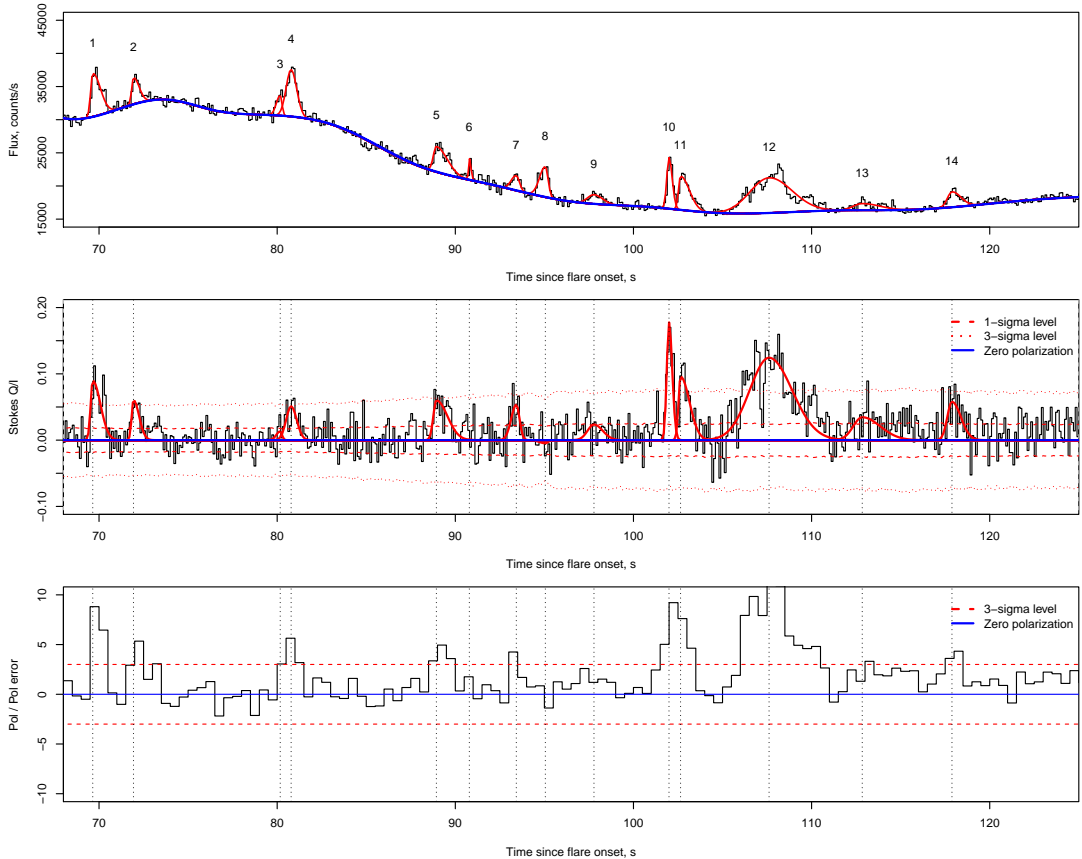}
    }}
  \caption{
    The region of UV Ceti flare near its maximum (see Figure~\ref{fig2}) with
    0.1 s resolution (upper panel) and the normalized Stokes $Q/I$ parameter
    with the same resolution (middle panel). Red lines are approximations of
    spikes with all parameters except amplitudes fixed to values listed in
    Table~\ref{table1}.
    The lower panel is the same quantity rebinned to 0.5 s resolution and
    normalized to its Poissonian errors.
    \label{fig5}}
\end{figure*}

\begin{figure}
  {\centering \resizebox*{1.0\columnwidth}{!}{
      \includegraphics[angle=270]{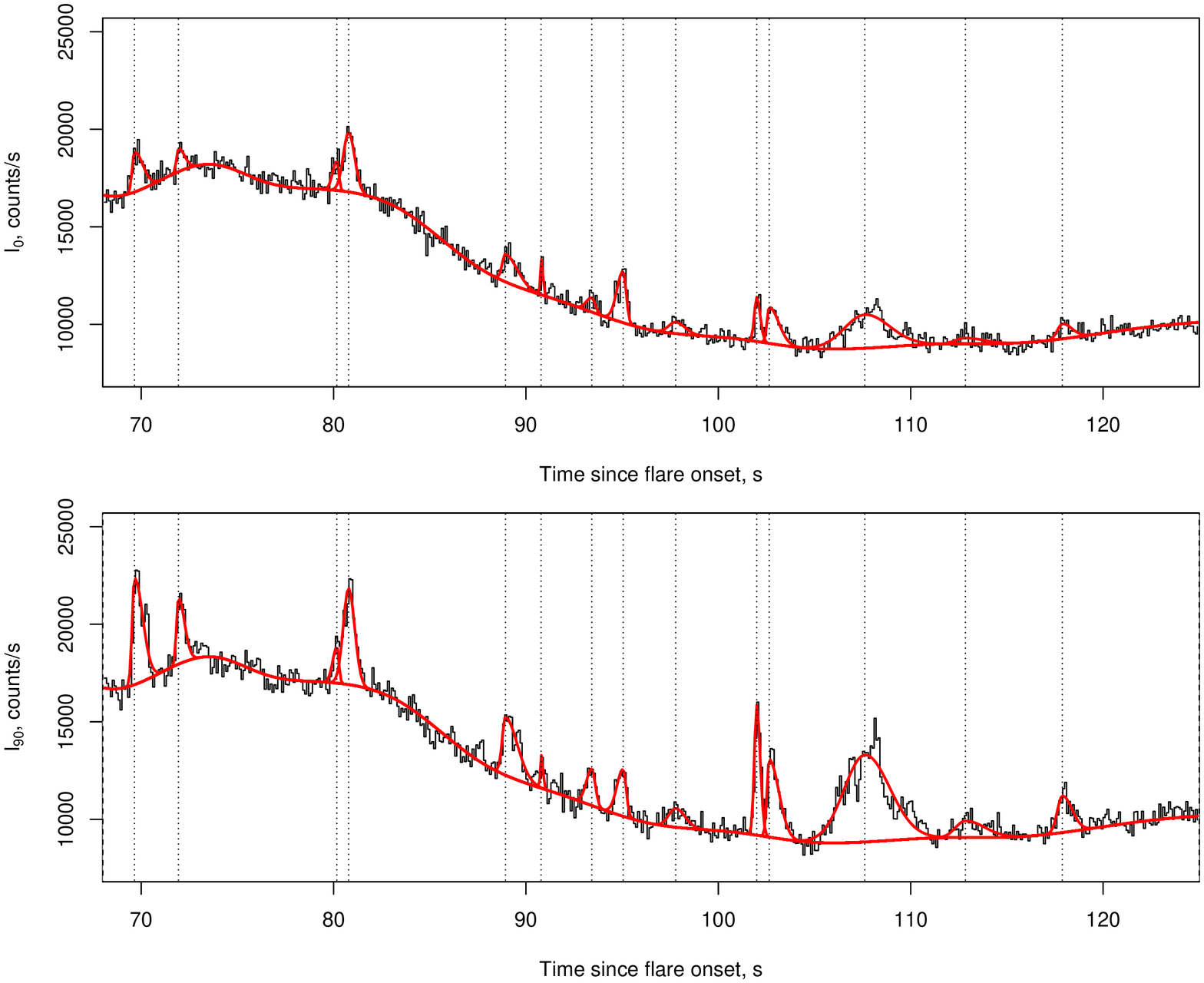}
    }}
  \caption{
    Independent fits of $I_0$ and $I_{90}$ polarized emission components of the
    region shown in Figure~\ref{fig5} with splitted Gaussians with all
    parameters except amplitudes fixed to values listed in Table~\ref{table1}.
    \label{fig5_1}}
\end{figure}

Figure~\ref{fig4} shows the temporal behaviour of $Q/I$ normalized Stokes
parameter, i.e. the ``projection'' of linear polarization onto the Wollaston
prism axis, over the 10 minutes of pre-flare and flare peak intervals. There
are no signs of any significant deviation from zero level ($p$-value of
Shapiro-Wilk normality test is $p=0.4$) except for the moments coinciding with
the spikes seen in the light curve. Figure~\ref{fig5} shows the close-up of the
interval containing these spikes with polarization fitted with the same
splitted Gaussian profiles with all the parameters fixed to values listed in
Table~\ref{table1} except for the amplitudes. It is clear that the positional and
morphological coincidence of spikes in intensity and normalized Stokes $Q/I$
parameter is perfect, while the amplitudes of the latters are in most cases
exceeding $2.5\sigma$ (spikes \textnumero\textnumero 1, 2, 4, 5, 7, 10, 11, 12, and 14). For a more
evident picture of their significances, the lower panel of Figure~\ref{fig5} shows the
same quantity as the middle one, but constructed from the light curves rebinned
to 0.5 s temporal resolution, so that the spikes fall into 1 to 3 bins, and the
RMS decreases by 2.23 times. Then the significance of polarized spikes is
better than $10^{-4}$, and the probability of overall effect to be random is
definitely lower than $10^{-20}$.

The polarization is absent anywhere except the spikes, therefore we may suggest
that the spikes represent an additional, polarized emission component superimposed
with the overall unpolarized giant flare. Then, the observed intensities may be
represented as a sum of ``flare'' and ``spike'' components as

\begin{equation}
  I_{0}(t) = I_{0}^{\rm flare}(t) + I_{0}^{\rm spike}(t)
\end{equation}
\begin{equation}
  I_{90}(t) = I_{90}^{\rm flare}(t) + I_{90}^{\rm spike}(t)
\end{equation}

Next, we may introduce an \textit{intrinsic} spike polarization, analogous to Eq.~\ref{eq_lower_limit}, but
corresponding to the polarization of a spike emission alone, excluding the
background flare emission:

\begin{equation}
  P_{\rm low,int}(t) = |Q^{\rm spike}(t)|/I^{\rm spike}(t) = \frac{|I_{0}^{\rm spike}(t)-I_{90}^{\rm spike}(t)|}{I_{0}^{\rm spike}(t)+I_{90}^{\rm spike}(t)}
  \label{eq_lower_limit_intrinsic}
\end{equation}

Figure~\ref{fig5_1} shows the $I_0$ and $I_{90}$ components over the
  spikes interval. The shapes in both components are quite similar, and nearly
  the same as spike shapes in total intensity, which may suggest that the
  polarization of every spike is more or less constant over time. Therefore we may
  characterize the \textit{mean intrinsic} polarization of every spike by
  independently fitting the $I_0$ and $I_{90}$ intensity profiles with splitted
  Gaussians described by Eq.~\eqref{eq_spike_shape} with all parameters except
  for the amplitude $A$ fixed to ones from Table~\ref{table1} (see
  Figure~\ref{fig5_1}) and substitute the corresponding amplitudes $A_{0}$ and
  $A_{90}$ into Eq.~\ref{eq_lower_limit_intrinsic} as mean intensities:

\begin{equation}
\left<P_{\rm low,int}\right> = \frac{|A_0-A_{90}|}{A_0+A_{90}} \ \ {\rm .}
\label{eq_pint}
\end{equation}

\begin{figure}
  {\centering \resizebox*{1.0\columnwidth}{!}{
      \includegraphics[angle=270]{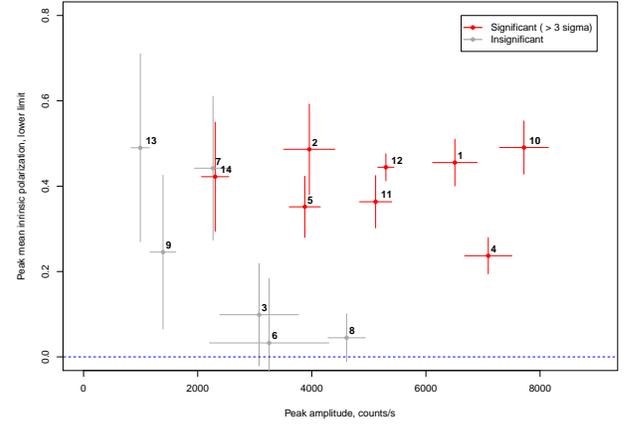}
    }}
  \caption{
    Lower limits on the degree of \textit{mean intrinsic} linear polarization of
    the spikes, computed according to Eq.~\ref{eq_pint}, versus the peak fluxes.
    Spikes with low significance polarization (less than $3\sigma$) are in gray,
    spikes with significant polarization (with significance levels $10^{-3} - 10^{-5}$) -- in red.
    \label{fig6}}
\end{figure}

Table~\ref{table1} and Figure~\ref{fig6} show these values along with corresponding errors. For six
spikes (\textnumero\textnumero 3, 6, 7, 8, 9, 13) the polarization lower limits
do not differ significantly from zero, exceeding it for less than one $\sigma$ (in
three cases) and for 1.5 -- 2 $\sigma$ (three more cases). The polarization of
eight other spikes is significant and quite high -- it peaks at 20\%
(\textnumero 4), 30\% (\textnumero\textnumero 5, 11), 40\%
(\textnumero\textnumero 1, 12, 14) and 45\% (\textnumero\textnumero 2, 10) and is
not correlated with either spike intensity or its duration.

\section{Discussion}

\subsection{Large linear polarization of spikes -- the evidence of their
  synchrotron origin}

In our observations of a giant UV Ceti flare on Dec 28, 2012 we discovered 14
spike bursts near its maximum, and clearly detected their polarization. Their
short durations, sufficiently high intensities and comparable scales of rise
and fading phases, and most importantly -- their high intrinsic polarization
exceeding 30-45\% -- suggest that these events may only be caused by the
synchrotron emission of ultra-relativistic electron streams moving in magnetic
fields of the corona.

Indeed, no other mechanism of emission generation and/or
transformation in astrophysical conditions can produce such level of linear
polarization.  For Thompson scattering its degree can't exceed 10--20\%
\citep{angel_1969,brown_1977}.  Linear polarization of bremsstrahlung radiation
is typically in 5--25\% range
\citep{brown_1972,emslie_2008}. 
Inverse Compton scattering on the electron beam does not change the state of
unpolarized emission, while for the intrinsic synchrotron (synchrotron-self
Compton mechanism) emission it lowers the original polarization degree by
several times -- its maximal degree is lower than 30-35\%
\citep{bonometto_1973,krawczynski_2012}. Therefore, the linear polarization
exceeding 35--45\% that we detected in sub-second spikes is a direct
proof of their synchrotron origin.

The synchrotron mechanism for optical flaring emission from red dwarfs had been
initially proposed by \citet{gordon_1954} who noted that the detection of
polarized flares might be the proof for it. However, the sparse and irregular (see
Introduction) polarimetric observations of UV Cet type stars have been
fruitless until now
\citep{karpen_1977, eritsyan_1978, tuominen_1989, alekseev_1994}. It seems that
the majority of flares, especially the longer ones studied in these works, are
dominated by thermal emission. Shorter ones are, however, still may be at least
partially driven by synchrotron emission, as our observations strongly suggest.


Below we will discuss possible origin of accelerated electrons, estimate their
energies and number densities, and show that they may indeed be naturally formed in UV
Ceti corona.



\subsection{On possible origin of ultra-relativistic electrons}

The multi-wavelength observations of the Sun and solar flares, as well as
active stars, which have analogous nature of flaring activity
\citep{gershberg_2015}, theoretical analysis and numerical simulations have
recently made it clear that the structure of magnetic fields here
is a complex system of small-scale magnetic knots, multiple organized thin
loops and regular thread-like structures with minimal scales about $10^8$ --
$10^9$ cm or even smaller, and with the magnetic field strength nearly constant along the
threads from corona to photosphere
\citep{shibata_2002, lopes_2008, meyer_2013, klimchuk_2015}.

During the magnetic reconnection in the corona, the collisionless Hall current
sheets, which may spontaneously form in a critical self-organizing state and
trigger the flaring energy release \citep{cassak_2008}, are being fragmented to
separate filaments due to tearing instabilities, and the particles are being
accelerated in these small-scale filaments or between them on a short time
scales \citep{drake_2006, che_2011}.


Possible mechanisms of electron acceleration up to 300--400 MeV energy
include super-Dreicer electric fields in magnetic reconnection regions what may
form the streams of ultrarelativistic electrons
\citep{craig_2002, gordovskyy_2010}. \citet{gordovskyy_2010} demonstrated by an
illustrative estimation that in the conditions of solar corona ($B \approx 100$
Gauss) the electrons may accelerate up to energies of several tens of MeV,
while having quite flat ($\delta \approx 1.5$) energetic spectrum. As the
strength of accelerating electric field $E \propto B^2$, for typical UV Ceti
coronal magnetic field of 300--1500 Gauss \citep{mullan_2006} the electrons may
reach the energies of hundreds of MeV. At the same time, \citet{craig_2002} and
\citet{litvinenko_2006} demonstrated that during the magnetic reconnections the
super-Dreicer electric field also forms and may accelerate the electrons up to
few hundred MeV -- several GeV with similar flat energetic spectrum.
The same electron energies may result from interaction of electrons with fast
magnetohydrodynamical modes during the acceleration \citep{yan_2008}.

The streams of ultrarelativistic electrons, formed due to acceleration
of background thermal particles, subsequently lose their energy radiatively
during the motion in a slowly-changing magnetic fields of separate threads,
producing the spike bursts we observed in UV Ceti flare.
Due to stochastic nature of particle acceleration the pitch angle
  distribution of ejected electrons is isotropic
  \citep{dalla_2005,minoshima_2008}, and their emission is therefore
  omnidirectional. Pitch-angle diffusion due to elastic scattering on whistlers
  \citep{stepanov_2007} or other fast MHD modes \citep{yan_2008} both keeps it
  isotropic and prevents the particles from streaming rapidly along the field
  lines, keeping them close to the acceleration region during the cool-down.
  Moreover, if the particle source is powerful enough the strong pitch-angle
  diffusion regime is realized and a the turbulent ``wall'' is formed when a
  cloud of high-energy particles propagates along the magnetic field with the
  velocity of about the phase velocity of waves which is much less than
  particle velocity \citep{bespalov_1991,trakhtengerts_2008}.
  Therefore the geometric effects of emission beaming and finite
  propagation time may be neglected in the analysis of the spikes.

Note that the giant flare itself (which seems to be is purely thermal)
and spike bursts are produced at different regions -- the former at a loop
footpoint in chromosphere (see \citet{gershberg_2015} and references therein),
while the latters -- in the corona itself. Moreover, these phenomena may in
principle be produced on two different stars (as both UV Cet and BL Cet are
flaring stars). Of course, the probability of latter case is quite small, but
it can not be completely ignored.

\subsection{What are the parameters of magnetic fields and electrons responsible for the spikes?}


If the spike bursts are indeed caused by synchrotron radiation, we may estimate
the range of magnetic field strengths, Lorentz factors and numbers of
accelerated electrons necessary to provide their obsreved peak luminosities and
fading durations, while keeping the emission in $U$ band. For simplicity, and
following the arguments presented earlier, we will assume the isotropic
distribution of electron pitch angles, and will use corresponding angle-averaged
formulae for their synchrotron emission.



For a single electron, the peak frequency of its synchrotron emission is
\citep{ginzburg_1965, rybicki_lightman_2004}
\begin{equation}
  \label{eq1}
  \nu_s \approx 1.2\times10^6 B \gamma^2 \ \ \mbox{Hz,}
\end{equation}
where $\gamma = E_s/mc^2$ is the Lorentz factor, $E_s$ -- the electron energy
and $B$ -- the magnetic field strength, while the characteristic timescale of
its energy loss is \citep{ginzburg_1965}
\begin{equation}
  \label{eq2}
  \tau_s \approx 5\times10^8 B^{-2} \gamma^{-1} \ \ \mbox{s.}
\end{equation}
By combining Eq.~\ref{eq1} and Eq.~\ref{eq2}, for the effective
frequency of $U$ band $\nu_s = 8\times10^{14}$ Hz  we may get
\begin{equation}
  \label{eq3}
  \gamma \approx 700 \left(\frac{\tau_s}{0.38\ \mbox{s}}\right)^{1/3}
\end{equation}
and
\begin{equation}
  \label{eq4}
  B \approx 1.4\times10^3\left(\frac{\tau_s}{0.38\ \mbox{s}}\right)^{-2/3} \ \ \mbox{Gauss.}
\end{equation}

For the average fading time of spikes (see Table~\ref{table1} and Section~\ref{sec_durations}) $\tau_s = 0.38$ s, from
Eq.~\ref{eq3} and Eq.~\ref{eq4} we get $\gamma\approx700$, which corresponds to
the electron energy $E_s = \gamma mc^2 \approx 360$ MeV, and
$B \approx 1.4\times10^3$ Gauss.  The latter value is in good agreement with the
magnetic field strength derived from observations of UV Ceti flares
\citep{mullan_2006, zaitsev_2008}.

The number of ultra-relativistic electrons
$N$ responsible for the synchrotron emission of the spike burst may then be
estimated as
\begin{equation}
  \label{eq5}
  N \approx \frac{W}{L}\ \ \mbox{,}
\end{equation}
where $W$ is its the observed luminosity in $U$ band and the
luminosity of a single electron is \citep{ginzburg_1965}
\begin{equation}
  \label{eq6}
  L \approx 1.6\times10^{-15} B^2 \gamma^2 \ \ \mbox{erg~s$^{-1}$.}
\end{equation}
Using
the average luminosity of spikes
$W=4.6\times10^{27}$ erg~s$^{-1}$ (see Table~\ref{table1}), $B = 1400$ Gauss and
$\gamma = 700$ we get for the number of emitting particles $N=3\times10^{30}$.

The scatter of actual spikes' fading times from $0.11$ s to $0.7$ s
(excluding two longer ones, see Table~\ref{table1}), as well as their luminosities, gives the ranges of $\gamma=460-860$,
$B=900-2300$ Gauss, $E_s=235-440$ MeV and $N = 4\times10^{29} - 8\times10^{30}$, correspondingly.

\subsection{The spectrum of accelerated electrons}

Now we may estimate the slope $\delta$ of electron energy distribution which is
necessary to have $N \approx 4\times10^{29} - 8\times10^{30}$ particles
accelerated up to energies exceeding $E_s \approx 440$ MeV to generate observed
spike bursts.

For simplicity we may assume that the accelerated electrons follow the same
power-law, $dN = N(E)dE \propto E^{-\delta} dE$, in
the wide range of energies from tens of keV up to hundreds of MeVs, which is
consistent with observations of solar (see, for example, \citet{ramaty_1994,
  kanbach_1993, lin_2011}) and stellar flares \citep{smith_2005}.



The lower energy cutoff $E_0$ for different solar flares have been found to be
10 -- 50 keV \citep{sui_2007,kontar_2008,caspi_2010}. For the analysis of
synchronous radio and X-ray observations of flaring stars the value of 10 keV
has been used \citep{smith_2005}.
The maximal energy $E_m$ of accelerated electrons may be as high as
  0.5-1 GeV, as both observations of solar flares
  \citep{ramaty_1994, kanbach_1993, lin_2011} and theoretical models
  \citep{craig_2002,litvinenko_2006,yan_2008} suggest.

Then the total number of accelerated particles is
\begin{equation}
  \label{eq_N0}
  N_0 = \int\limits_{E_0}^{E_m}N(E)dE \propto \frac{E_0^{1-\delta}-E_{m}^{1-\delta}}{\delta-1}
  \mbox{,}
\end{equation}
and the mean energy of accelerated particle
\begin{equation}
  \label{eq_Emean}
  \left<E\right> = \frac{1}{N_0} \int\limits_{E_0}^{E_m} E N(E) dE =
  \frac{\delta-1}{\delta-2} \frac{E_0^{2-\delta}-E_{m}^{2-\delta}}{E_0^{1-\delta}-E_{m}^{1-\delta}}
  \mbox{.}
\end{equation}

All the particles with energies exceeding $E_s$ sooner or later contribute to $U$
band emission during the spike burst (as for the higher energy electrons
cooling time is shorter, according to Eq.~\ref{eq2}), therefore the number of
electrons responsible for the synchrotron emission of the spike burst is
\begin{equation}
  \label{eq_N}
  N = \int\limits_{E_s}^{E_m}N(E)dE = N_0 \frac{E_s^{1-\delta}-E_{m}^{1-\delta}}{E_0^{1-\delta}-E_{m}^{1-\delta}}
  \mbox{.}
\end{equation}

The number of accelerated electrons may also be written as
\begin{equation}
  \label{eq_nt}
  N_0 = k n_t V
  \mbox{,}
\end{equation}
where $k$ is the fraction of thermal electrons with density $n_t$ being accelerated during the
magnetic reconnection in the filament zone with characteristic size $l$ and
volume $V=\pi l^3/6$ (assuming spherical shape).
As an upper estimate of $l$ we will use the minimal scale of
inhomogeneities in solar (or stellar) corona of $10^8-10^9$ cm
\citep{shibata_2002, meyer_2013}. 
The efficiency of thermal particles acceleration $k$ in solar flares
it may reach 10\%--100\% (see for example \citet{lin_2011}). On the other hand,
for flares on UV Cet type stars it is not determined yet, and we will consider
below the interval $k = 10^{-7} - 1$.
Finally, for the thermal density $n_t$ on UV Ceti 
we will assume the interval from $10^{10}$ cm$^{-3}$ in a quiet state
\citep{gudel_2009} to $10^{12}$ cm$^{-3}$ during the flares
\citep{mullan_2006}.  

As the energy for particle acceleration is originating from the magnetic field,
we may write the (probably over-conservative) energy budget condition as
\begin{equation}
  \label{eq_E}
  \frac{\left<E\right> N_0}{V} = \left<E\right> n_t k < \frac{B^2}{8\pi}
  \mbox{.}
\end{equation}

\begin{figure}
  {\centering \resizebox*{1.0\columnwidth}{!}{
      \includegraphics[angle=0]{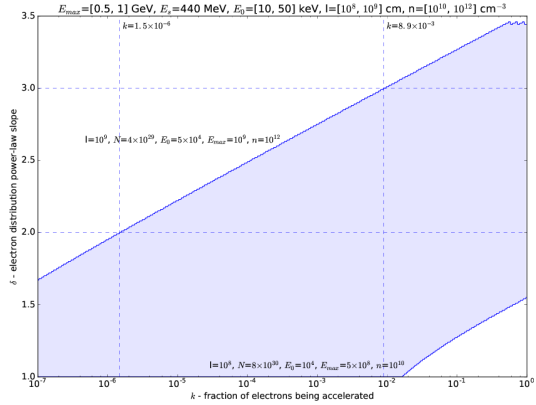}
    }}
  \caption{
    Allowed power-law slopes $\delta$ of electron distribution and fractions
    $k$ of thermal electrons being accelerated,
    necessary to
      generate the observed spike bursts for different parameter values.
      \label{fig7}}
\end{figure}

Then we may solve Eqs.~\ref{eq_Emean}, \ref{eq_N}, \ref{eq_nt} and \ref{eq_E} for possible
values of $\delta$ on $k$ allowed
for the aforementioned set of parameters: $E_0=10-50$ keV, $E_m=0.5-1$ GeV,
$N=4\times10^{29}-8\times10^{30}$, $n_t=10^{10}-10^{12}$ cm$^{-3}$ and $l=10^8-10^9$
cm. The result is shown as a filled region in Figure~\ref{fig7}
bounded by the lines with parameters marked there. One can readily see that
even for the lowest fraction of accelerated electrons $k=10^{-7}$ there are
always parameters of coronal plasma that may explain the generation of observed
spike bursts, if the electron spectral slope is steeper than
$\delta=1.7$. On the other hand, to have spectral slope flatter than $\delta=2$
and $\delta=3$, one have to accelerate more than $~10^{-4}$\% and $~1$\%
thermal particles, correspondingly. The former case is valid for any combination
of $E_0$, $l$ and $N$, while the latter -- just for some subset of it.
The upper limit on $\delta \lesssim 3.4$ is placed by the energy density
condition of Eq.~\ref{eq_E}.

It is clear that the observed spike bursts may be generated for any combination of
possible parameters of the corona if one have $1 < \delta < 3.4$ and
$k>10^{-7}$. The larger luminosities and/or smaller sizes of active regions
correspond to the harder spectra with smaller slopes, to get the necessary
amount of electrons with sufficiently large energies. Such values of $\delta$ differ
from typical slopes of energy spectrum of electrons in solar flares, typically
greater than 3 and reaching values of 4 -- 6 \citep{aschwanden_2006}. On the other
hands, there are evidences for the detection of quite flat spectra of electrons
with slopes of 2 -- 3 in the flares of UV Ceti type stars \citep{smith_2005}.


Therefore, we may conclude that the conditions for the synchrotron origin of
the detected highly polarized spike bursts may naturally occur in the corona of
UV Ceti, with no additional assumptions except for the contemporary views on coronal
activity of the Sun and flaring stars.

\section{Conclusions}

In our observations of a giant flare of UV Ceti on Dec 28, 2012 we discovered 14
spike bursts near its maximum, and clearly detected their linear polarization,
which intrinsic value exceeds 35\% -- 40\%. These events in such numbers, and
the polarization of the flare emission in general, have never been seen before
from any UV Ceti type star. We argue that their short durations, sufficiently
high intensities and comparable scales of rise and fading phases, and most
importantly -- the polarization, suggest that these events may only be caused by
the synchrotron emission of ultrarelativistic electrons moving in magnetic
fields of the corona. As we demonstrated in Section~4, the UV Ceti corona may
indeed possess the conditions -- densities and magnetic field strengths --
necessary to accelerate significant amount of particles up to the energies of
hundreds of MeV producing the emission in $U$ band with the durations about
fractions of a second. Therefore, our results is the first ever evidence for
the generation of ultrarelativistic electrons with such energies in the
coronae of UV Ceti type stars.

As a side note, let's also mention that the presence of very energetic
particles in flares of UV Cet type stars may have significant impact on the
planets orbiting them. Indeed, the fraction of such particles may be ejected
away from the star and may significantly worsen the conditions for the
appearance and development of life on the planets formally inside the habitable
zones around the red dwarfs \citep{kasting_1993, tarter_2007}. It is important
as such a stars are easier targets for exoplanet detection, as the recent
observations suggest \citep{guinan_2014, gillon_2014}.

\section*{Acknowledgements}

The work has been partially supported by RFBR grant \textnumero 15-02-0828 and Program
\textnumero 7 of Russian Academy of Sciences.
Observations were partially
carried out according to the Russian Government Program of Competitive Growth
of Kazan Federal University. The theoretical analysis of the possible influence
of red dwarfs activity to habitable planets was performed under the financial
support of the grant of Russian Science Foundation \textnumero 14-50-00043.

\nocite*{}
\bibliographystyle{pasa-mnras}
\bibliography{uvcet}

\end{document}